\documentstyle[12pt]{article}
\begin{document}
\newcount\sectionnumber
\sectionnumber=0

\title{CP Violation and Leptogenesis}
\author{Andy Acker${}^{a)}$, Hisashi Kikuchi${}^{a)}$, Ernest Ma${}^{a)}$ and
Utpal Sarkar${}^{b)}$}
\date{\mbox{}}

\maketitle
\begin{center}
${}^{a)}$ {\it Physics Department, University of California,
Riverside, CA 92521, USA.}\\

${}^{b)}$ {\it Theory Group, Physical Research Laboratory,
Ahmedabad - 380009, India.}\\

\end{center}

\begin{abstract}

Recently a model of chaotic inflation was proposed, where the
right handed sneutrinos drive the baryogenesis. We study some of
the details of the model, particularly the aspect of CP
violation, and determine the number of right handed sneutrinos required
for the viability of such models.

\end{abstract}

\vspace{1 in}
\begin{center}
UCRHEP-T109
\end{center}

\newpage

Recently a model has been proposed \cite{model}, where the inflation is
driven by the superpartner of the right handed neutrino.
Although the inflationary scenario is quite appealing for the
solution of the horizon and the flatness problem \cite{hflat}, it is
difficult to formulate a particle physics model consistently.
In addition to solving this problem, this model generates
(B-L) asymmetry at lower temperature in the post inflationary
universe through violation of lepton number. This model requires
the right handed neutrinos which are singlets under all groups
for one of these fields to be the inflaton. The Majorana mass
terms for these fields have been introduced in this model by
hand to explain the COBE result \cite{COBE}, but otherwise there is no
natural reason for the origin of that scale in the model.

The main objective of this article is to
study the CP violation in these models,
determine the number of singlet fields required for CP
violation and count the number of CP-violating phases which can be
present in the combination of the Yukawa couplings appearing in
the processes under consideration \cite{cpviol}. We shall also give a
natural explanation for the origin of the Majorana mass scale.
We consider the particle content to be exactly same as that
of ref 1, with the exception of the number of right handed
neutrinos ($N_{\alpha}$'s). We
start with $\bar{n}$ number of these scalars ($N_{\alpha}, \:\; \alpha = 1,
... ,\bar{n}$) all of which carry a (B-L) number --1.
While any of the $N_{\alpha}$ can serve as the
inflaton, for the time being let us consider the scalar
partner of $N_1$ to be the inflaton. We do
not give explicit mass to these fields. The only mass scales in
this model are the grand unified symmetry breaking scale $M_u
\sim \langle \Sigma \rangle$, where $\Sigma$ is a 24-plet of
SU(5) and breaks the SU(5) symmetry, and $M_w \sim \langle H_u
\rangle \sim \langle H_d \rangle$, where $H_u$ and $H_d$ are
$5-$ and $\bar{5}$-plets of
SU(5) and mediates the standard model symmetry breaking and give
masses to the fermions. We also assume that the singlet fields
have odd R-parity like other fermions.

The Yukawa part of the superpotential is given by,
\begin{equation}
W = h^{\alpha j} N^c_{\alpha} l_j H_u + k^{ij} e_i^c l_j H_d
\end{equation}
where, $\alpha, \beta = 1, ... ,\bar{n}$ are the indices for the
fields $N_\alpha$ and $i,j = 1, ... ,n$ are the indices for the
left handed leptons. We assume that (B-L) is a global symmetry of
the superpotential, hence higher order terms involving the
$N_{\alpha}$'s are forbidden.
Although finally we are interested in $n = 3$, for our analysis
we shall take $n$ to be arbitrary.

Until the time $H \sim M_{\alpha \beta}$ the
$\tilde{N}_1$ field will roll down the potential and after that
oscillate around its minimum, $\tilde{N}_1=0$, with frequency
$M$. Once the universe cools down to about $10^4$ GeV, the
coherent oscillation starts to decay into lighter particles thus
reheating the Universe again to about $10^{10}$ GeV. During this
period of reheating the decay of $\tilde{N}_1$ field produces
(B-L) number asymmetry through lepton number violation due to CP
violation. This (B-L) asymmetry remains as a baryon number asymmetry
after the electroweak anomalous processes \cite{anom}.

While the origin of the $N_{\alpha}$ mass term is not critical
to our discussion, we would like to suggest an interesting possibility.
After the field $\Sigma$ acquires a {\it vev}, there will be new
gravity induced nonrenormalizable contributions to the Yukawa
coupling, which are not protected by any symmetry. Such terms
have been considered in the literature to solve the bad fermion
mass relations in SU(5) GUT and some other contexts \cite{hidim}. We  are
interested in the term  which involves the fields $N_{\alpha}$, this
is given by,
\begin{equation}
W = \displaystyle \frac{1}{M_P} H_{\alpha \beta} N_\alpha
N_\beta \Sigma \Sigma.
\end{equation}
When the field $\Sigma$ acquire a {\it vev}, namely, $\langle
\Sigma \rangle \sim M_u \sim 10^{16}$ GeV (for supersymmetric
SU(5) GUT\cite{gutrev}), this term will induce a lepton number violating mass
term for the fields $N_i$ of the right magnitude $\sim
\frac{{M_u}^2}{M_P} H_{\alpha \beta} \sim M_{\alpha \beta} \sim
10^{13}$ GeV to be consistent with the COBE data \cite{mass}, $\delta T/T \sim
10^{-5}$.

In the generation of the (B-L) asymmetry through the lepton
number asymmetry \cite{model,anom,fy,luty}, the CP violation
plays a crucial role \cite{sak},
and we now study
systematically this question of CP violation and the decay of
the heavy neutrinos in a general way, trying to determine the number of
phases present in the Yukawa type couplings. The CP
violating parameter $\epsilon$, which is the asymmetry in the
$\tilde{N}_{\alpha}$ decay between that into leptons $l_i \tilde{H}_u$
and that into antisleptons $\tilde{l}^*_i H_u^*$,
is defined as
\begin{equation}
\epsilon=\frac{|A(\tilde{N}_{\alpha}\to l_i \tilde{H}_u)|^2 -
|A(\tilde{N}_{\alpha}\to \tilde{l}^*_i H^*_u)|^2}{|A|^2}
\end{equation}
and is given by (see Fig.1 of ref. \cite{model}),
\begin{equation}
\epsilon = \frac{\ln 2}{8 \pi} \frac{{\rm Im} (h_{\alpha i}
h_{\alpha j} h^*_{\beta i} h^*_{\beta j})}{h_{\alpha i}
h^*_{\alpha i}} = \frac{\ln 2}{8 \pi} \frac{T_{\alpha
\beta i j}}{|h_{\alpha i}|^2}
\end{equation}
with $t_{\alpha \beta i j}= (h_{\alpha i}
h_{\alpha j} h^*_{\beta i} h^*_{\beta j})$
and $T_{\alpha \beta i j} = {\rm Im} (t_{\alpha
\beta i j})$. While this expression is clearly invariant under
rephasing of the fields
$l_i$ \cite{luty}, the invariance of $\epsilon$ under the
transformation $N_\alpha \to e^{i \delta_\alpha} N_\alpha$ is not
obvious. Indeed, under such a transformation
the quantity $t_{\alpha \beta
i j} \to e^{2 i (\delta_\alpha - \delta_\beta)} t_{\alpha \beta
i j}$. It would  appear that a suitable choice of the phases $\delta_\alpha -
\delta_\beta$ can make $t_{\alpha \beta i j}$ real, and thus give
$\epsilon = 0$. However, rephasing of the fields
$N_\alpha$ introduces new phases in the propagator and wave function
through complex phases in the mass parameter
which exactly cancel the extra phase picked up by
$t_{\alpha \beta i j}$, leaving $\epsilon$ invariant.
In the leptogenesis models we need only consider the
effect of CP violation in the decay of heavy
neutrinos at temperatures well above the electro-weak symmetry breaking scale,
where the Higgs field has not acquired a $vev$, and the left handed neutrinos
are massless.
We thus choose to work in the basis
where $N_{\alpha}$ are the real, positive mass eigenstates of
the right handed neutrinos,
and note that this choice exhausts our freedom to make phase transformations
on the fields $N_{\alpha}$.

The CP violating parameter
describing the asymmetry in the decay of Majorana neutrinos
$N_{\alpha}$ into leptons
vs. antileptons is proportional to the same combination of Higgs-Yukawa
couplings $T_{\alpha \beta i j}$ appearing in Eq. 4 (see ref. \cite{luty}).
We wish to determine under what conditions $T$ is non-zero, and
in general, find the number of independent $T$'s which can appear in
the asymmetry parameter. From the definition of the $T$'s,
one can immediately get the following relations,
\begin{equation}
T_{\alpha \beta i j} = T_{\alpha \beta j i} = - T_{\beta \alpha
i j} = - T_{\beta \alpha j i}
\end{equation}
{\it i.e.,} the $T$ is antisymmetric under the interchange of
the heavy singlet indices, while it is symmetric under the
interchange of the $\nu_i$'s. Thus, for $\bar{n} = 1$, {\it
i.e.,} only one generation of heavy singlet field $N_\alpha$
there is no CP violation. This relation, however, does not
determine the minimum number neutrinos required for CP violation.

We now proceed to calculate the number of phases that can appear in
$T_{\alpha\beta ij}$ when $n$, the number of left handed neutrinos,
and $\bar{n}$, the number of right handed neutrinos, are arbitrary.
We find that the maximum number of phases which can occur is
equal to the number of independent phase in the
$\bar{n}\times\bar{n}$ hermitian matrix $H$
\begin{equation}
H_{\alpha\beta}=h_{\alpha j}h^*_{\beta j}=(hh^{\dagger})_{\alpha\beta},
\label{H}
\end{equation}
where $h_{\alpha j}$ is the Yukawa coupling between $N_{\alpha}$ and
$l_j$.
To see this recall that the left handed neutrinos are massless,
and hence degenerate. Therefore, there are no observable consequences to
performing a unitary transformation on the fields $l_i$;
in particular, the magnitude of CP violation in the decay of $N_{\alpha}$
must be invariant under such transformations. Inspection of Eq. 4
reveals that $H_{\alpha\beta}$ is the minimum set of Yukawa couplings
appearing in the CP violation parameter which is invariant
under this transformation. It should be noted that unitary transformations
of the $l_i$ leave the structure of the Standard Model lagrangian unchanged.

If $n \ge \bar{n}$ $H$ is an unconstrained $\bar{n}\times \bar{n}$ Hermitian
matrix, whereas if $n < \bar{n}$ there are additional constraints the
number of independent elements in $H$. We find that the number
of independent complex phases in $H$ is given by\footnote{This is the
same number of phases that can occur in an $\bar{n}\times\bar{n}$ Majorana
mass matrix\cite{cpviol}, however; the physical context here is quite
different.}
\begin{equation}
N_{{\rm phases}}= \left\{ \begin{array}{ll}
                   \frac{\bar{n}(\bar{n}-1)}{2}  & n \ge \bar{n}    \\
                  n\bar{n}-\frac{n(n+1)}{2}   & n < \bar{n}
                   \end{array} \right.
\end{equation}
For the case of $n\ge\bar{n}$ its possible to choose a basis where
the matrix of Higgs Yukawa couplings $h_{\alpha i}$ can itself be written as
an $\bar{n}\times \bar{n}$ hermitian matrix.
To see this explicitly note that for
$n \ge {\bar n}$ we can construct a unitary matrix $V_{jk}$
such that
\begin{equation}
h^{ \alpha j}V_{jk}=0\; {\rm for}\: k>{\bar n}.
\end{equation}
By performing the unitary transformation on the lepton doublet $l_j$,
$l_k=V^{-1}_{kj}l_j$ it is clear that
only the $l_k$ with $k\le{\bar n}$ couple to the right-handed singlet
fields $N_{\alpha}$.
In this basis the couplings are completely described by an
${\bar n} \times {\bar n}$ square matrix $h^{\prime}$.
Finally, recall that any square matrix $h^{\prime}$ can be written as the
product $h^{\prime} =hW$, where $W$ is unitary and h is hermitian.
Thus by again performing a unitary transformation on the lepton doublets
$l_k$,
the Higgs Yukawa coupling becomes
\begin{equation}
h^{\alpha j}N^c_{\alpha}l_jH_u
\end{equation}
where, as advertised, $h$ is an ${\bar n} \times {\bar n}$ hermitian
matrix.

We may use the fact that $h$ is hermitian to show
how the complex phases manifest themselves in the
quantity $T_{\alpha\beta i j}$, and do this explicitly
for $\bar{n}  =2 $ and $3$.
(If $\bar{n}=1$, $h$ is real and $T$ vanishes identically, hence,
as discussed above there is no CP violation).
For $\bar{n}=2$, $h$ contains one phase, and there are
three distinct $T$'s: $T_{1211}$, $T_{1212}$ and $T_{1222}$.
These are related to each other by
\begin{equation}
\frac{T_{1211}}{|t_{1211}|}=\frac{T_{1212}}{|t_{1212}|}=
\frac{T_{1222}}{|t_{1222}|}=\sin(2\delta_{12})
\end{equation}
where $\delta_{\alpha j}$ is defined by $h^{\alpha j}=|h^{\alpha j}
|e^{i\delta_{\alpha j}}$.
For $\bar{n}=3$ there are 18 distinct $T$'s, however, there are only 3
independent phases, appearing in 9 possible combinations.
These are summarized by the following three relations
\begin{equation}
\frac{T_{\alpha\beta\alpha\alpha}}{|t_{\alpha\beta\alpha\alpha}|}=
\frac{T_{\alpha\beta\beta\beta}}{|t_{\alpha\beta\beta\beta}|}=
\frac{T_{\alpha\beta\alpha\beta}}{|t_{\alpha\beta\alpha\beta}|}=
\sin(2\delta_{\alpha\beta})
\end{equation}
\begin{equation}
\frac{T_{\alpha\beta\alpha\gamma}}{|t_{\alpha\beta\alpha\gamma}|}=
\frac{T_{\alpha\beta\beta\gamma}}{|t_{\alpha\beta\beta\gamma}|}=
\sin(\delta_{\alpha\beta}+\delta_{\alpha\gamma}-\delta_{\beta\gamma})
\end{equation}
\begin{equation}
\frac{T_{\alpha\beta\gamma\gamma}}{|t_{\alpha\beta\gamma\gamma}|}=
\sin(2\delta_{\alpha\gamma}-2\delta_{\beta\gamma})
\end{equation}
where $\alpha < \beta$ and $\gamma \ne \alpha \ne \beta$.
In this notation the three independent phases appearing in
$T_{\alpha\beta i j}$ are
$\delta_{12}$, $\delta_{13}$ and $\delta_{23}$.

In conclusion we have examined
leptogenesis models where the
initial lepton number asymmetry is generated by a CP
asymmetry in the decay of a heavy neutrino(sneutrino) into
leptons with respect to the decay into anti-leptons(sleptons).
The Higgs Yukawa couplings $h_{\alpha i}$ that gives rise
to these decays must contain at least one complex phase.
We have calculated the number of complex
phases that can occur in this coupling for an arbitrary number of heavy
right handed neutrinos  and massless left handed neutrinos.
We find that, for $\bar{n}$ heavy right handed neutrinos
and $n$ left handed neutrinos the number of
independent phases appearing in the asymmetry
parameter $\epsilon$ is given by equation (7).
Thus a minimum of 2 right handed neutrinos(sneutrinos) and one left handed
neutrino are required for CP violation to occur in right handed neutrino decay.
For the 3 generation case ($n=\bar{n}=3$) there are three independent
phases in the Higgs Yukawa couplings. Finally, we note that this result is
qualitatively different from CP violation in charged-current weak interactions
arising from a complex phase in the Kobayashi-Maskawa matrix.

\begin{center}\subsection*{Acknowledgements}\end{center}
Two of us (U.S. and A.A.) wish to thank the Department of Physics at the
University of California, Riverside for their hospitality.
A.A. also wishes to thank the Australian Department of Immigration
for making possible the trip to UCR. This work
was supported in part by Department of Energy contract number
DE-AT03-87ER40327.

\eject

\end{document}